# Gradient bandgap enables >13% efficiency sulfide Kesterite solar cells with open-circuit voltage over 800 mV

Kang Yin[1,3]†, Jinlin Wang[1,3]†, Licheng Lou[1,3]†, Xiao Xu[1,3], Bowen Zhang[1,3], Menghan Jiao[1,3], Jiangjian Shi[1]*, Dongmei Li[1,3,4], Huijue Wu[1], Yanhong Luo[1,3,4]* and Qingbo Meng[1,2,3,4]*

[1]Beijing National Laboratory for Condensed Matter Physics, Renewable Energy Laboratory, Institute of Physics, Chinese Academy of Sciences, Beijing, 100190, P. R. China.

[2]Center of Materials Science and Optoelectronics Engineering, University of Chinese Academy of Sciences, Beijing, 100049, P. R. China.

[3]School of Physical Sciences, University of Chinese Academy of Sciences, Beijing, 100049, P. R. China.

[4]Songshan Lake Materials Laboratory, Dongguan, 523808, P. R. China.

*Corresponding author. Email: qbmeng@iphy.ac.cn; shijj@iphy.ac.cn; yhluo@iphy.ac.cn.

†These authors contributed equally to this work.

**Abstract:**

Sulfide Kesterite $Cu_2ZnSnS_4$ (CZTS), a nontoxic and low-cost photovoltaic material, has always being facing severe charge recombination and poor carrier transport, resulting in the cell efficiency record stagnating around 11% for years. Gradient bandgap is a promising approach to relieve these issues, however, has not been effectively realized in Kesterite solar cells due to the challenges in controlling the gradient distribution of alloying elements at high temperatures. Herein, targeting at the Cd alloyed CZTS, we propose a pre-crystallization strategy to reduce the intense vertical mass transport and Cd rapid diffusion in the film growth process, thereby realizing front Cd-gradient CZTS absorber. The Cd-gradient CZTS absorber, exhibiting downward bending conduction band structure, has significantly enhanced the minority carrier transport and additionally improved band alignment and interface property of CZTS/CdS heterojunction. Ultimately, we have achieved a champion total-area efficiency of 13.5% (active-area efficiency: 14.1%) in the cell and in particular a high open-circuit voltage of >800 mV. We have also achieved a certified total-area cell efficiency of 13.16%, realizing a substantial step forward for the pure sulfide Kesterite solar cell.

Kesterite $Cu_2ZnSn(S, Se)_4$ (CZTSSe) thin-film solar cells have attracted extensive attention in recent years due to their environmental friendliness, low cost, and high stability[1-5]. In particular, Se-based Kesterite cells have achieved power conversion efficiency (PCE) close to 15%[6], demonstrating promising application of these material systems. Pure-sulfide $Cu_2ZnSnS_4$ (CZTS) is an important branch of Kesterite materials. The bandgap of CZTS can be easily adjusted within a wide range of 1.3 to 2.1 eV through alloying of metal elements (Cd, Ge or Ag)[7-10]. This tunability makes CZTS highly suitable not only for single-junction solar cells but also for the uppermost cells for tandem cell stacks[3]. However, despite numerous research efforts were paid in recent years[3,7,11,12], the photoelectric performance of sulfide Kesterite cells still faces major bottlenecks, with certified efficiencies remaining stagnate at around 11% for several years[3,13]. The main limitation lies in the low open-circuit voltage ($V_{OC}$)[14-17], with the highest reported value only reaching ~63% of its theoretical limit ($V_{OC}^{SQ}$) until now[9].

Theoretical studies indicate that deep-level defects in CZTS materials, such as $Sn_{Zn}$ and $V_S$-$Cu_{Zn}$, have extremely large electron capture cross-sections ($10^{-12}$ $cm^2$) due to the strong electron-phonon coupling.[16,18] In addition, compared to CZTSe, CZTS has a higher electron effective mass and thus lower charge transport ability[19,20]. Moreover, the high-temperature sulfurization process for CZTS often introduces significant Cu/Zn disorder, also inducing the formation of high concentration deep-level defects[15,21-23]. The presence of Cu/Zn disorder and deep-level defects leads to energy band and potential fluctuations, which further increases charge transport scattering and diminishes charge transport performance[24,25]. These intrinsic semiconductor properties of CZTS materials pose challenges for charge transport and defect control, consequently limiting further enhancement of device performance.

The construction of gradient conduction band structure within the absorber has been considered as a promising approach to improve transport ability of minority carriers over a large spatial scale, thus minimizing non-radiative charge recombination losses[26-28]. The gradient bandgap strategy has obtained wide success in $Cu(In,Ga)Se_2$ (CIGS) solar cells[26,27] but still encounters challenges in Kesterite solar cells, despite of extensive research efforts. For example, Wu et al. attempted to fabricate an Ag gradient-alloyed CZTSSe film[29]. However, Ag diffusion and gradient distribution is difficult to control at high temperatures and Ag alloying is not effective in controlling the conduction band structure[30]. Researchers have also explored introducing S/Se gradient into the kesterite absorber[31-33]. While different melting points and reaction activities of Se and S pose challenge in achieving precise control of the S/Se ratio and spatial distribution during the complex selenization (sulfurization) process[4,31,34,35]. Cd gradient has also been observed in Kesterite cells. It was found that heterojunction annealing can induce Zn/Cd interdiffusion, resulting in a slight Cd gradient on the surface of the Kesterite absorber[36]. Although Cd alloying can lower the conduction band of Kesterite[3,7,30], this heterojunction annealing induced gradient distribution is insufficient to construct a gradient bandgap at a submicron scale. Due to the aforementioned reasons, constructing a gradient conduction band within CZTS absorber remains a significant challenge in the field.

In this work, we proposed a pre-crystallization strategy involving Cd alloying to successfully achieve a gradient bandgap engineering for solution-processed CZTS. The pre-crystallization strategy on one hand induces the homogeneous nucleation within CZTS film and thus reduces the intense vertical mass transport during film growth process, and on the other hand improves the initial crystal quality of CZTS, thus increasing the diffusion barriers of Cd. These collective effects render a significant Cd gradient be realized in the CZTS absorber. The Cd-graded CZTS exhibited a conduction band-dominated gradient bandgap, having effectively promoted the carrier transport. Additionally, the Cd-rich CZTS surface has also improved the conduction band alignment of kesterite/CdS and reduced the interface defect. Ultimately, we have achieved a champion efficiency of 13.5% (active-area efficiency: 14.1%) in the Cd-gradient CZTS cell, and in particularly realized a high $V_{OC}$ of >800 mV, whose $V_{OC}/V_{OC}^{SQ}$ value (67.9%) significantly surpasses all other reported results[9,37]. Through the gradient Cd, we have also achieved a certified total-area efficiency of 13.16%, representing a substantial step forward for the pure sulfide Kesterite solar cell.

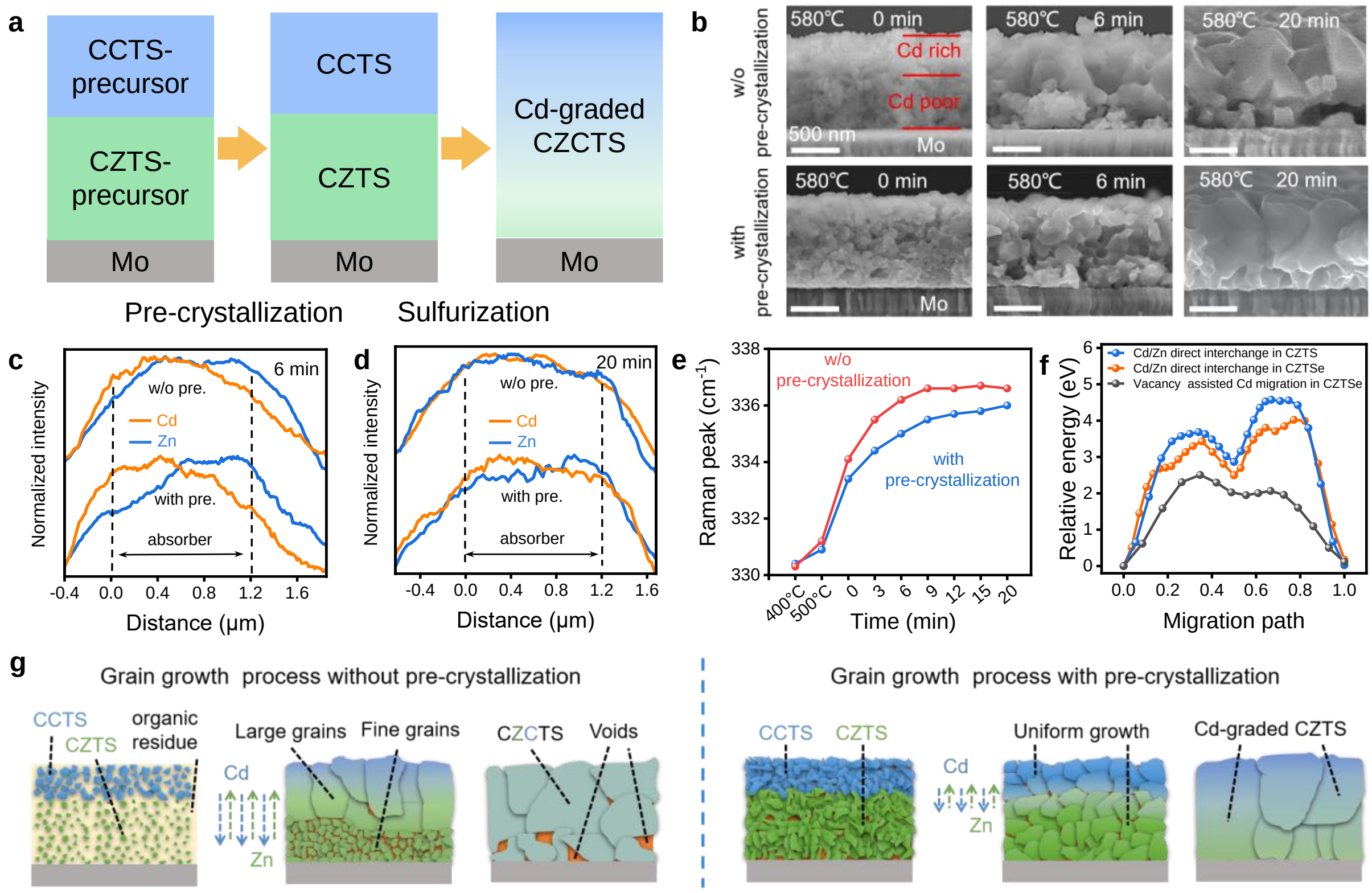

**Fig. 1. Influence of pre-crystallization on the sulfurization process.** a. Schematic diagram of the process to prepare the Cd-graded CZTS absorber through pre-crystallization and sulfurization. b. SEM images of samples (with or without pre-crystallization) at different sulfurization stages (580 °C for 0 min, 6 min, 20 min). c-d. Cross-sectional EDX elemental profile (normalized by the maximum value) of these samples. e. Raman peak evolution of CZTS phase of these two samples at different sulfurization stages (532 nm excitation). f. Calculated relative energy of Zn/Cd interchange as a function of migration path through direct or vacancy-assisted mechanism. g. Schematic diagram of the grain growth process of the film without (left) or with (right) the pre-crystallization process.

## Pre-crystallization strategy to form Cd-gradient CZTS absorber

The Cd as an alloying element in pure sulfide CZTS materials shows great potential for achieving a gradient bandgap. Firstly, the bandgap of Cd alloyed CZTS changes linearly with the Cd/(Cd+Zn) ratio ranging from 0 to 0.4[38-40]. Secondly, Cd alloyed CZTS can effectively suppress Cu/Zn disorder and improve crystal quality, resulting in good photoelectric performance over a wide range of Cd substitution[3,8,10,12,41,42]. Thirdly, the incorporation of Cd in CZTS can theoretically regulate the conduction band position, thus enabling the realization of a conduction band gradient design and modifying the unfavourable conduction band energy alignment of CZTS/CdS heterojunction[3,7,30]. Based on these foundations, the primary focus of our work is to achieve a gradient distribution of Cd in the CZTS.

In our experimental approach, precursor solutions of CZTS and $Cu_2CdSnS_4$ (CCTS) were prepared separately for the sequential deposition CZTS and CCTS precursor layers onto the Mo substrate, obtaining a Mo/CZTS/CCTS precursor film. To address the issue related to rapid interdiffusion of Zn/Cd at high-temperature sulfurization, we proposed a strategy of introducing pre-crystallization (as depicted in Fig. 1a). This pre-crystallization was carried out by pre-annealing the dual-layer precursor film at 420 °C for 10 minutes under $H_2S$ (10%)/Ar atmosphere (Fig. 1a, Supplementary Note 1 and Supplementary Fig. 1). To elucidate the influence of pre-crystallization on the sulfurization process, cross-sectional scanning electron microscope (SEM) images and corresponding energy dispersive X-ray (EDX) elemental mapping of the films with or without pre-crystallization were captured at different sulfurization stages (Fig. 1b-d). It can be seen that the sample without pre-crystallization exhibited preferential crystallization in the top region, where upper grains rapidly merged into larger crystals and engulfed fine crystals beneath them, accompanied by a rapid interdiffusion of Zn/Cd. This process led to the formation of numerous voids at the bottom of the final absorber, with no obvious gradient distribution of Zn/Cd. In contrast, the pre-crystallized samples exhibited uniform nucleation and crystallization throughout the film. Along with the slower diffusion of Zn/Cd, well-arranged large crystals with a gradient distribution of Cd were ultimately formed.

Raman measurement was employed to characterize the influence of elemental interdiffusion on the CZTS phase formation process. Due to the interdiffusion induced elemental composition change, Raman spectra, especially the peak position of Kesterite phase, exhibited obvious evolution during the sulfurization process (Supplementary Fig. 2). Specifically, Raman peaks of both the samples blue shifted continuously as sulfurization proceeded (Fig. 1e), indicating the gradual upward diffusion of Zn towards the CCTS layer. It can be further seen that the Raman peak of the sample without pre-crystallization showed a rapid shift in the initial stage and kept almost constant after 9 minutes of sulfurization, indicating a completion of Zn/Cd diffusion. Comparatively, Raman peaks of pre-crystallized sample exhibited a slower shift and did not reach equilibrium even the sulfurization ended. For the final state, the pre-crystallized sample possessed lower-wavenumber Raman peak than that of the control sample, indicating higher Cd content. These results demonstrated that the pre-crystallization has effectively impeded the interdiffusion of Zn/Cd elements, thus enabling the realization of element gradient. The inhibitory effect of pre-crystallization on elemental diffusion is relatively straightforward to understand. In the absence of pre-crystallization, the film usually undergoes surface nucleation and top-down growth behavior, similar to the selenization crystallization process[4,34]. This growth mode is accompanied by a rapid upward diffusion of elements at the onset of reaction, resulting in local elemental disparities, lattice distortion and a substantial generation of atomic vacancies. In contrast, pre-crystallized samples, owing to the prior removal of organic residues and the initiation of homogeneous nucleation and crystallization in advance, exhibit uniform growth process, leading to a reduction in vacancies and a diminished driving force for elemental interdiffusion. In this case, the exchange of Zn/Cd mainly transpires at the CCTS/CZTS interface with a gentler process. To provide further clarity, we employed DFT theoretical calculations to evaluate the energy barriers for Zn/Cd interchange through direct or vacancy assisted mechanism. The results (Fig. 1f) firstly indicated that, compared to CZTSe,

Cd possesses a larger diffusion barrier in CZTS, owing to smaller lattice sizes. The results also suggested that to sustain the high diffusion barrier, vacancy path in the lattice need be eliminated, which thus highlighted the significance of enhancing the quality of the initial crystals in the film as realized by the pre-crystallization process. For clarity, these two different grain growth processes are schematically illustrated in Fig. 1g.

Notably, besides the gradient, other positive effects that caused by the Cd alloying itself, such as promoting crystallization, inhibiting the formation of ZnS secondary phase, minimizing Sn loss during sulfurization, and reducing Cu/Zn disorder, were also observed in our experiment (Supplementary Fig. 3-4).

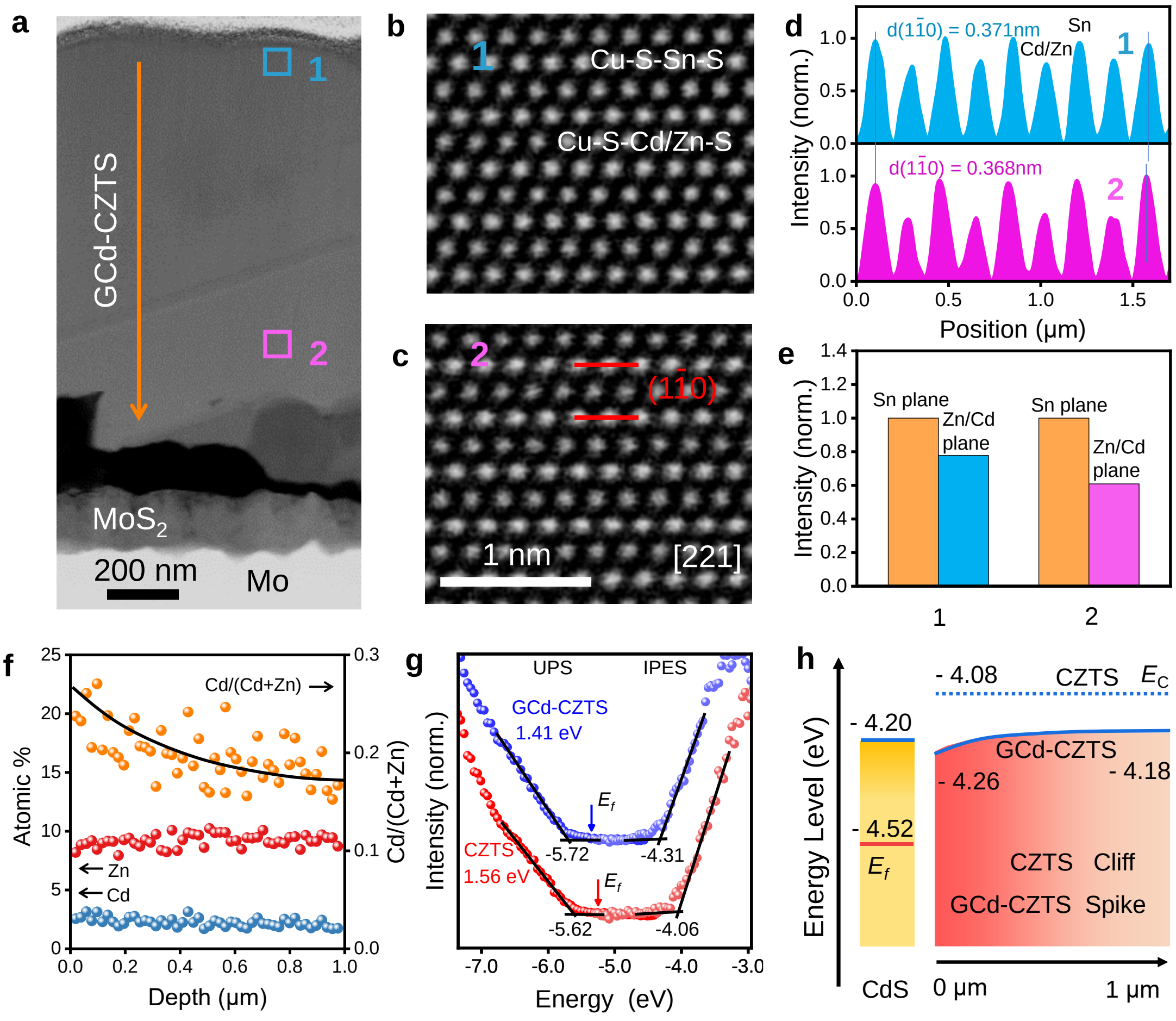

**Fig. 2. Elemental and energy band distribution.** a. Cross-sectional HAADF STEM image of the GCd-CZTS sample. b-c. Filtered HAADF atomic images taken from region 1 and region 2 in the film. d. Averaged intensity profiles of ($1\bar{1}0$) plane and e. normalized total intensity of Sn

plane and Zn/Cd plane of the atomic images of region 1 and region 2. f. STEM-EDX elemental distribution of Zn and Cd in the film (along the orange line as marked in the STEM image). g. UPS and IPES spectra of CZTS and GCd-CZTS absorbers. h. Energy band distribution of the GCd-CZTS sample. The band-edge positions were collectively evaluated according to the element distribution, photoelectron spectroscopy and Kelvin-probe force microscopy results.

**Gradient elemental composition and energy band characteristics**

We used scanning transmission electron microscopy (STEM) and high-angle annular dark field (HAADF) imaging to further quantitatively investigate the distribution of Zn/Cd in the Cd-gradient CZTS (GCd-CZTS) absorber. As in Fig. 2a, the cross-sectional STEM image exhibited uniform HAADF contrast across the entire film with negligible grain boundaries. The film also showed clear selected-area electron diffraction patterns corresponding to the [221] observation direction of the Kesterite lattice (Supplementary Fig. 5a). These results demonstrated a high crystallization quality of the fabricated GCd-CZTS absorber. The HAADF atomic images of the film were further captured (Fig. 2b-c), which showed the same atomic arrangements in the top and bottom regions. Along the $[1\bar{1}0]$ direction, alternate arrangements of light and dark contrast corresponding to Cu-Sn and Cu-Zn/Cd planes can be clearly observed (Supplementary Fig. 5b). In Fig. 5d-e, quantification of the averaged HAADF contract profiles revealed the $(1\bar{1}0)$ interplanar spacing of the region 1 is 0.371 nm, which is slightly larger than that of the region 2. Moreover, when using the HAADF intensity of the Cu-Sn plane as a reference, the Cu-Zn/Cd plane of the region 1 showed an obviously higher intensity that of the region 2, indicating more Cd in the Kesterite lattice of the top region.

The Cd/Zn distribution in the film was further determined using STEM-EDX measurement. As in Fig. 2f, the Cd/(Cd+Zn) ratio exhibited a gradient distribution across the absorber film, exceeding 0.26 in the surface region and gradually reducing to ~0.16 to the back interface. According to linear relationship between the Cd/(Cd+Zn) ratio and the bandgap of Cd alloyed

CZTS materials, the bandgap of GCd-CZTS in the surface and bottom regions was evaluated to be 1.41 eV and 1.47 eV, respectively (Supplementary Fig. 6a-b). This bandgap of the surface region of the absorber film was also measured by ultraviolet photoelectron spectroscopy (UPS) and inverse photoelectron spectroscopy (IPES), which also demonstrated a bandgap of 1.41 eV (Fig. 2g). Compared to the pure CZTS, the downward shift of the conduction band minimum reached 250 meV. UPS and Kelvin probe force microscopy (KPFM) methods were further used to evaluate the band edge position of the Cd-alloyed CZTS with 0.26 or 0.16 Cd/(Cd+Zn) ratio (Supplementary Fig. 6). It revealed that the gradient in the conduction band between the surface and the bottom absorber reached ~80 meV, as schematically shown in Fig. 2h, which should be able to facilitate the electron transport. According to this evaluated band structure, it can also be found that a spike energy band alignment was formed at the heterojunction interface, which would help reduce the interfacial charge recombination.

**Photoelectric characterization of film and device**

We further characterized the influence of gradient band structure on the carrier electron transport characteristics of the absorber using emission wavelength-dependent time-resolved photoluminescence (PL) spectroscopy (Fig. 3a-c)[43]. For clarity, the time for the PL maxima at different wavelengths is marked out by black lines and the transient PL probed at short and long wavelength are extracted (880nm and 975nm for CZTS, 920nm and 1025nm for GCd-CZTS). For the GCd-CZTS, its emission at 920 nm exhibited a sharp drop behavior in the early stage, while the emission at 1025 nm rose slowly. A time delay of about 4 ns was observed between these two emission bands. The matching of the dynamics between the 920 nm-emission decay and the concomitant rise of the 1025-nm emission implied the possibility of ultrafast carrier transfer between these two emission states. This phenomenon strongly supported the formation of gradient band structure and its driven carrier transfer within the absorber. Comparatively, no such behaviour was observed in the gradient-free CZTS sample.

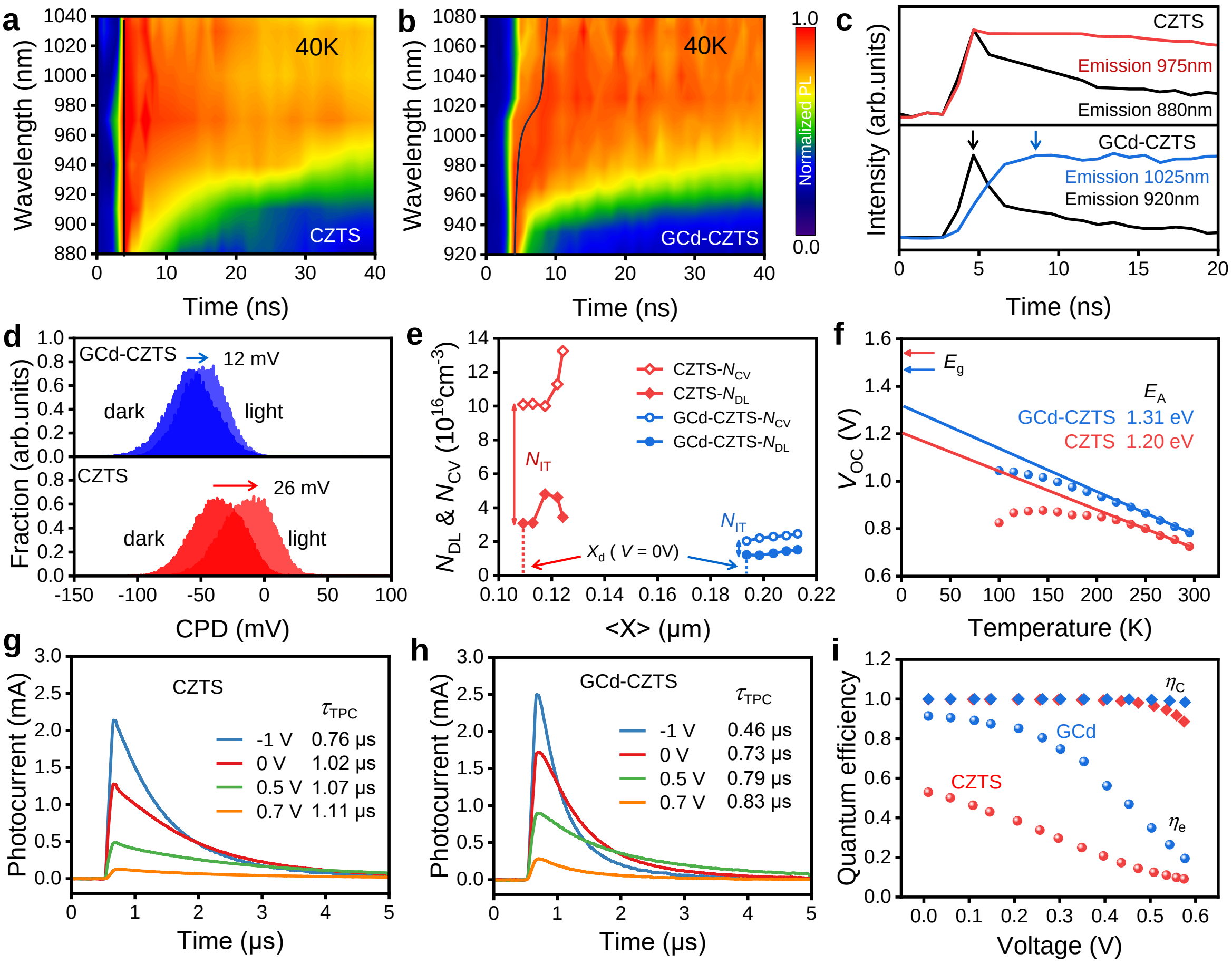

**Fig. 3. Photoelectric characterization of film and device.** a-b. Two dimensional pseudocolor plots of transient PL as functions of emission wavelength and time delay for CZTS and GCd-CZTS films (40 K, excitation at 640 nm). The black line marks the time of the PL maximum at different wavelengths. c. Time-resolved PL decay at different emission wavelengths (CZTS: 880 nm and 975 nm, GCd-CZTS: 920 nm and 1025 nm). The arrows show the time delay of different emissions of GCd-CZTS sample. d. Contact potential difference (CPD) distribution of CZTS and GCd-CZTS samples in the dark and under illumination. e. Charge profiles of the CZTS and GCd-CZTS cells obtained by driven level capacitance profiling ($N_{DL}$) and capacitance-voltage ($N_{CV}$) methods (measured at 3 kHz). f, Temperature-dependent $V_{OC}$ of the devices. Red and blue arrows refer to the bandgap of CZTS and GCd-CZTS, respectively. g-h. Photocurrent decay behaviors of the cells under various bias voltages ($\tau_{TPC}$, exponential decay time). i. Charge collection ($\eta_c$) and extraction ($\eta_e$) efficiencies of the cells derived from modulated electrical transient measurements.

In addition to confirming the fast carrier transfer, we also investigated surface electric and defect properties of these films using KPFM. As in Fig. 3d, CPD distribution of the GCd-CZTS sample was little influenced by the measurement condition (dark and illumination), indicating negligible surface defect and charge trapping behaviors. Comparatively, the averaged CPD of the pure CZTS sample shifted by about 26 mV, due to the defect charge trapping induced energy band bending[4]. These results were further supported by a direct measurement of charge distribution within the absorbers using drive-level capacitance profiling (DLCP) and capacitance-voltage (*C*-*V*) methods[44] (Fig. 3e). By calculating the charge density difference between these two methods, the interfacial defects density of the GCd-CZTS sample ($8.1\times10^{15}$ $cm^{-3}$) was found to be nearly an order of magnitude lower than that of the CZTS sample ($6.9\times10^{16}$ $cm^{-3}$). Additionally, GCd-CZTS sample showed much lower bulk charge density, indicating a reduction in the Cu/Zn substitution defects through Cd alloying. This consequently increased the depletion width of the absorber, which would facilitate the carrier transport in a larger spatial region. Due to these benefits, in the temperature-dependent $V_{OC}$ measurement the difference between the charge recombination energy barrier ($E_A$) and the bandgap of the fabricated GCd-CZTS cell was much smaller than that of the CZTS solar cell[45] (Fig. 3f).

We further used modulated transient photocurrent to investigate the overall charge transport and charge loss characteristics of the fabricated cells (Fig. 3g-i). With the improved energy band structure and facilitated carrier transport, the GCd-CZTS device demonstrated much smaller photocurrent decay time. In addition, photocurrent peak intensity of this device also exhibited smaller decline compared to that of the control CZTS device as the bias voltage increased, implying more efficient and more stable carrier extraction from the absorber to the buffer/window layers. Together with the photovoltage decay properties (Supplementary Fig. 7), the charge extraction ($\eta_e$) and collection ($\eta_C$) efficiencies that can reflect the charge loss in the bulk and interface regions of the cells have been quantified[46,47]. It is apparent that the GCd-CZTS cell possessed much higher $\eta_e$ in the entire voltage range, *i.e.* much lower bulk charge loss, primarily benefited from the enhanced carrier transport. This cell also showed higher $\eta_C$ at high voltage of >0.5 V, *i.e.* lower charge loss caused by interface charge recombination.

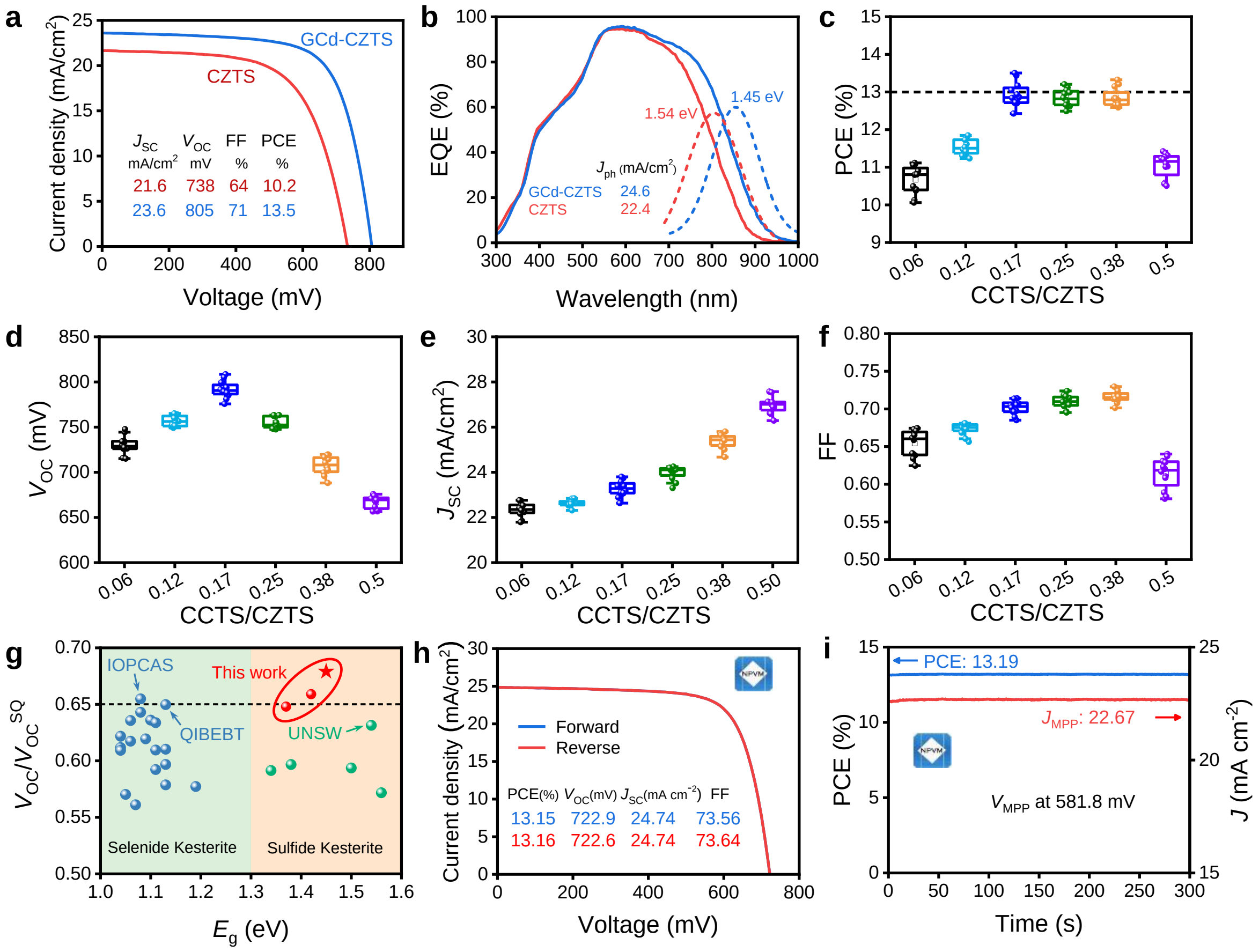

**Fig. 4. Performance of photovoltaic devices.** a. Current-voltage (*I-V*) characteristics of champion GCd-CZTS and control CZTS solar cells. b. EQE spectra of the cells. The bandgap was determined as the position of the maxima of 1st derivative of the EQE spectra. c-f. Statistical results of the performance parameters of GCd-CZTS devices with different CCTS/CZTS ratios. g. A summary of $V_{OC}/V_{OC}^{SQ}$ values of representative Kesterite solar cells that have been reported. h. Certified *I-V* curves and i. steady-state output performance of a GCd-CZTS device fabricated with CCTS/CCTS = 0.38.

## Device performance

For the solar cell fabrication, we have further optimized CCTS/CZTS ratio of the precursor film and finally achieved an impressive photoelectric conversion efficiency (PCE) of 13.5% (active-area PCE: 14.1%) at CCTS/CZTS ratio of 0.17 (Supplementary Fig. 8), with short-circuit current density ($J_{SC}$) of 23.6 mA $cm^{-2}$, fill factor (FF) of 0.71, and $V_{OC}$ of 805 mV (Fig. 4a). The PCE improvement compared to the control CZTS device exceeded 30% and moreover the $V_{OC}$ has reached the highest value reported so far for the sulfide Kesterite solar cells with bandgap <1.6 eV[9,48]. As shown in Fig. 4b, external quantum efficiency (EQE) spectra agreed well with the cell current-voltage characteristics and additionally revealed that this GCd-CZTS absorber had an averaged bandgap of 1.45 eV, a little smaller than that of the pure CZTS (1.54 eV). By further comparing the solar cells fabricated from different precursors, we found that high PCE higher than 13% can all be obtained in a wide range of the CCTS/CZTS ratios ranging from 0.17 to 0.38 although these cells had different performance parameters (Fig. 4c-f). Specifically, at 0.17 ratio, the cell had the highest $V_{OC}$ while at 0.38, the cell exhibited higher $J_{SC}$ and FF. Overall, these cells using the GCd-CZTS absorber all had superior performance than using the uniform Cd alloying, which only obtained ~12% PCE even after a systematic composition optimization (Supplementary Fig. 9). Our experiments further demonstrated that for the cells without using the pre-crystallization strategy, even if a gradient precursor film was used, comparative cell PCE also cannot be realized (Supplementary Fig. 1a). These results thus highlight the significance of the pre-crystallization realized gradient bandgap in realizing high performance sulfide Kesterite solar cells.

We further investigated the voltage deficits of these cells by comparing the cell $V_{OC}$ with their theoretical limits[18]. As in Fig. 4g, the highest $V_{OC}/V_{OC}^{SQ}$ of our cell reached 67.9%, obviously surpassing the latest reported 63.2% result obtained in the Ge-alloyed and gradient-free CZTS solar cell[9] (detailed data can be found in Supplementary Fig. 10 and Table 1). Our result is even higher than that obtained in the CZTSSe solar cells with PCE of >14.5%[49,50], thus demonstrating the significant effect of the gradient bandgap in reducing the voltage loss of Kesterite solar cells. One of our cells fabricated with CCTS/CZTS ratio at 0.38 was also sent out to an accredited independent laboratory (National PV Industry Measurement and Testing Center, NPVM) for certification. As in Fig. 4h, the certified total-area PCE reached 13.16% (certified test report is shown in Supplementary Fig. 11). Furthermore, when kept at the maximum power point at 581.8 mV, the cell gave a constant current output of 22.67 mA $cm^{-2}$, confirming a steady-state PCE of 13.19%. Compared to previously reported certified results[13], our cell here has achieved a significant efficiency enhancement, thus representing a substantial step forward for the pure sulfide Kesterite solar cell.

**Conclusion**

In this study, we have developed a gradient energy band structure within the Cd-alloyed CZTS absorber to facilitate bulk carrier transport so as to reduce charge and $V_{OC}$ loss of the sulfide Kesterite solar cell. In particular, we explored a pre-crystallization strategy that overcomes the persistent challenge in the construction of gradient bandgap in CZTS films, namely rapid diffusion of elements. Specifically, this strategy achieved uniform synchronous nucleation and crystallization at different spatial locations of the absorber layer film and improved the crystal quality, thereby significantly slowing down the Zn/Cd element interdiffusion during the high-temperature sulfurization reaction and realizing a gradient element distribution in the absorber. The Cd/Zn distribution introduced obvious gradient in the conduction band of the absorber, which one hand significantly facilitated the bulk carrier transfer and one the other hand improved the energy band alignment and defects at the heterojunction interface. As a result, the Cd-gradient CZTS solar cell demonstrated a high total-area PCE of 13.5% with impressively high $V_{OC}$ of 805 mV. Moreover, we also achieved a certified total-area efficiency of 13.16% in the Cd-gradient CZTS cell, which exhibited a significant enhancement compared to previous results, thus representing a substantial step forward for the pure sulfide Kesterite solar cell.

**Acknowledgements**

Authors thank Mr. Wentao Wang of IOP for STEM support and Professor Zhenghua Su from Shenzhen University for his valuable discussions and insightful suggestions to this work. This work was supported by the National Natural Science Foundation of China (Grant nos. 52222212 (J. S.), U2002216 (Q. M.), 52172261 (Y. L.), 52227803 (Q. M.), 51972332 (H.W.)). J. S. also gratefully acknowledges the support from the Youth Innovation Promotion Association of the Chinese Academy of Sciences (2022006).

**Author contributions**

Kang Yin, Jiangjian Shi, Yanhong Luo and Qingbo Meng conceived the idea and designed the experiments. Kang Yin, Jinlin Wang and Licheng Lou did the experiments and the data analysis. Licheng Lou, Xiao Xu, Bowen Zhang and Menghan Jiao supported CZTS solar cells fabrication. Huijue Wu and Dongmei Li supported M-TPC/TPV characterization and discussions. Kang Yin, Jiangjian Shi, Yanhong Luo and Qingbo Meng participated in writing the manuscript.